# SEMANTIC VISUALIZATION AND NAVIGATION IN TEXTUAL CORPUS

Férihane Kboubi, Anja Habacha Chaibi and Mohamed BenAhmed

RIADI-ENSI, University Campus of Manouba
2010, Manouba, Tunisie
Ferihane.Kboubi@riadi.rnu.tn, Anja.Habacha@ensi.rnu.tn,
Mohamed.benAhmed@riadi.rnu.tn

## ABSTRACT

*This paper gives a survey of related work on the information visualization domain and study the real integration of the cartography paradigms in actual information search systems. Based on this study, we propose a semantic visualization and navigation approach which offer to users three search modes: precise search, connotative search and thematic search. The objective is to propose to the users of an information search system, new interaction paradigms which support the semantic aspect of the considered information space and guide users in their searches by assisting them to locate their interest center and to improve serendipity.*

## 1. INTRODUCTION

Available information on Internet grows at an exponential rate. Data in these information systems is becoming more complex and more dynamic. As users with different backgrounds, traits, abilities, dispositions, and intentions increase dramatically, users' needs also become more diverse and complicated. Therefore the demand for a more effective and efficient means for managing and exploring data became a pressing issue. This poses a challenge to the traditional approaches and techniques used in current information retrieval systems. These systems use a keyword-based search process which is discontinuous because users have no control over the internal matching process which is not transparent to users. Besides, the output of search systems as result list presentation is linear and has a limited display capacity. Relationships and connections among documents are rarely illustrated. The retrieval environment lacks an interactive mechanism for users to browse. These inherent weaknesses of traditional information retrieval systems prevent them from coping with the sheer complexity of information needs and the multitude of data dimensionality.

The query-based search engines support only one search type "the precise search" which supposes that user know exactly for what they look for: a precise paper knowing its title, authors and major theme). It is not unusual for users to input search terms that are different from index terms used by the system. It will be very interesting to offer to users other search type such as "thematic search" (allowing users to navigate in the corpus according to a particular theme), "connotative search" (allowing users to discover the associated and similar concepts of their interest concepts) or "exploratory search" (allowing users to make an idea about the content of the corpus; and after a preliminary consultation that they will exactly define their needs).

As regards to the visualization methods, the study carried out by [1] showed that the result lists return an enormous quantity of information. This leads to a cognitive overload for users who cannot, in the majority of the cases, consult all the returned documents.

An innovative idea to guide users in their searches is to provide them an interaction method allowing them locating their needs throw the navigation in the document informational space. This type of interaction benefits from an important characteristic of the human cognition: it is easier to the users to discover or to locate for what they look, than to produce formal descriptions of information which they do not have. So, navigation within maps can replace advantageously writing of queries as far as semantics, being more explicit in maps, limits the problems of confusion and ambiguity often met in the query-based systems. Based on this innovative idea, the goal of this work is to find and propose solutions for these evocated problems.

The remaining of this paper is organized as following. In section 2, we present a survey of existing semantic cartography paradigms and discuss about their incorporation on information retrieval process. In section 3, we describe our semantic visualization approach which supports three search types: precise search, thematic search and connotative search. In this section we present two navigation approaches. The first one is based on domain ontology and the second is based on association relations.

## 2. SEMANTIC CARTOGRAPHY

Information retrieval visualization refers to a process that transforms the invisible abstract data and their semantic relationships in a data collection into a visible display and visualizes the internal retrieval processes for users. Basically, information retrieval visualization is comprised of two components: visual information presentation and visual information retrieval.

According to Card [2] and Tricot [3] there are three visual information presentation paradigms (which they called also cartography paradigms):

- The *representation paradigms*. They allow representing the structure of information. We distinguish between five types of information structures which are: the tabular structure [4], the treelike structure [5], the graph structure (Hypergraph [23] and TouchGraph systems), the temporal structure (ThemeRiver [6], spiral representation [7]) and the agglomerative structure (Themescapes [8]).

- The *visualization paradigms*. They represent the means of displaying information representations in a clear and coherent way on a limited space so that a person can become aware quickly of the presented information. Visualization techniques are classified in two groups: uniform visualization techniques (overview+details [9]) and the not uniform visualization techniques (document lens [10], the elusive walls [11], fisheye [12]).

- The *interaction paradigms*. They concern techniques allowing users to interact with the produced visualizations like: zoom and pan, focus and context, dynamic filtering [13], semantic zoom [14][15].

The visual information presentation provides a platform where visual information retrieval is performed or conducted. According to Zhang there are three information retrieval visualization paradigms [16]:

- The *QB* paradigm (Query searching and Browsing). Initially a query is required to limit the set of search results. Then a visualization of these results is constructed in which users may browse to concentrate their visual space for more specific information.

- The *BQ* paradigm (Browsing and Query searching). A visual presentation of the information set is first established for browsing. Then users submit their search queries to the visualization environment and corresponding search results are highlighted or presented within the visual presentation contexts.

- The *BO* paradigm (Browsing Only). This paradigm does not integrate any query searching components.

For a more detailed survey on semantic cartography paradigms see [17]. In spite of the variety of cartography paradigms proposed in the literature, their concrete integration on web remains however very limited and this for two main considerations. In the first place, from the user point of view, numerous are the ones who are not familiarized yet with these new paradigms. Secondly, as regards to material and software configurations, a big part of equipments connected on the net are not adapted to this type of applications.

Nevertheless, the evolution of hardware performance and the considerable development in the domain of interactive information visualization for years, allowed the emergence of new systems integrating information visualization techniques with varied levels, such as: Kartoo (http://www.kartoo.com), Toolnet (http://www.toolenet.com), Ujiko (http://www.Ujiko.com) and ArnetMiner (http://www.arnetminer.org).

All these systems are based on query definition as a search mode and they offer to users a graphical result maps as output. However, interaction means given to users remain elementary (selection, zoom). There are no means of semantic interaction and navigation in the informational space.

Based only on a query search mode, these systems support only a single search type which is the precise search (where users know exactly for what they look for). It would be very useful to propose to users other search mode guiding and assisting them in their searches and allowing them to navigate in the produced maps to refine their searches and to discover new knowledge.

## 3. OUR SEMANTIC MULTIFACET NAVIGATION APPROACH

The principal idea of this work is to propose a model allowing to put in evidence semantic inherent to the textual corpus. Our model is based on the result of the semantic annotation and indexation of textual documents [18][19][20]; and represents a new model of graphic visualization and semantic navigation (Figure 1).

The annotation process generates three types of annotations: descriptive annotations, conceptual annotations and thematic annotations.

−   The *descriptive annotations* are relative to: bibliographic annotations (title, authors, publication date), content descriptors (author abstract and key-words), technical annotations (format, size).

−   The *conceptual annotations* are relative to the concepts evoked in the document, their respective pertinence degree, and their respective association relations.

−   The *thematic annotations* are relative to the major theme and the set of minor themes handled in the document, and to the thematic association relations.

Our information representation and visualization model is based on the cartography paradigms studied in the state of the art. The aim is to reduce the cognitive effort of readers as regards to the classical result list representation mode. Indeed, graphical visualizations allow putting in evidence the pertinent information for users. According to Gershon and Page [21] the visualization amplifies the cognition and it allows to users and readers to observe, to understand and to make sense of these information.

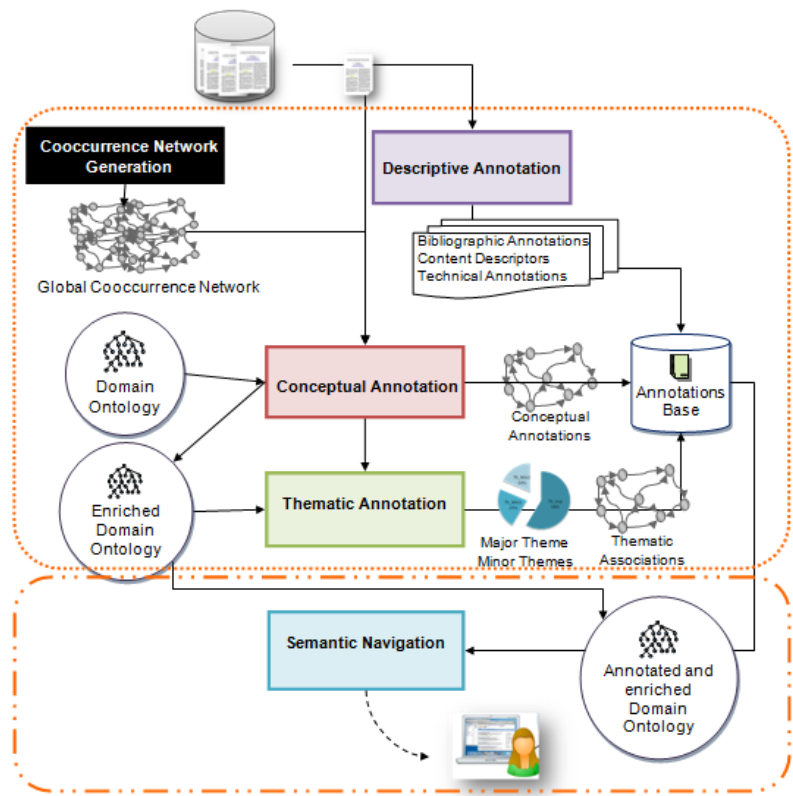

Figure 1. General principle of our explorative and thematic search approach

For representing and visualizing the information, we used a graph shaped representation based on the fisheye visualization techniques. This type of representation is adequate for representing semantic relations in the annotated domain ontology and the association networks (hierarchical relations between the concepts, the association relation between the concepts, the similarity relations between the documents, etc.). The fisheye technique allows putting in evidence the interest center of the user when he navigates in the graph.

In order to experiment our interactive visualization scenarios, we used the hypergraph-0.6.3 applet[1] since it is based on graph representation and fisheye visualization paradigms.
Our new interaction mode offers to users a multi-approach of semantic navigation:

- *Domain ontology based navigation approach* allowing users to make thematic searches to explore document informational space according to their themes of interest.

- *Concept association based navigation approach* allowing the users to make connotative searches by navigating in the conceptual association graphs.

- *Similarity relation based navigation approach* allowing users to make another type of connotative searches by navigating in the document similarity relation graphs.

### 3.1. Navigation guided by the domain ontology

The idea is to visualize the semantic content of the textual document corpus through a graphic representation of the annotated domain ontology. Initially the domain ontology is visualized as a hierarchy of themes and concepts, in which a user can navigate from one theme to another and from one concept to another in order to localize his interest center (Figure 2). For a given concept, the user can ask to display the titles of all documents indexed by this concept and to order them by their pertinence degree. The user can afterward consult the description of a document of his choice. This description represents a semantic summary of the selected document and contains descriptive, conceptual and thematic annotations already extracted during the annotation step.

Several advantages ensue from this navigation approach. Effectively, this navigation approach offers a thematic search mode by reflecting for a given domain the semantic common to the majority of users. It offers to users a representation of knowledge close to the cognitive model which they have on the domain, what avoids them getting lost in the semantic map and allows them to localize quickly their interest center.

Nielsen [22] came up with three fundamental questions that the users (Internet surfers) face when they navigate the cyberspace: *where am I now?, where have I been?, and where can I go next?*. This navigation approach helps users answer these questions and minimize the problems of lost in information space and disorientation syndrome.

---

[1] Available on line at http://hypergraph.sourceforge.net/

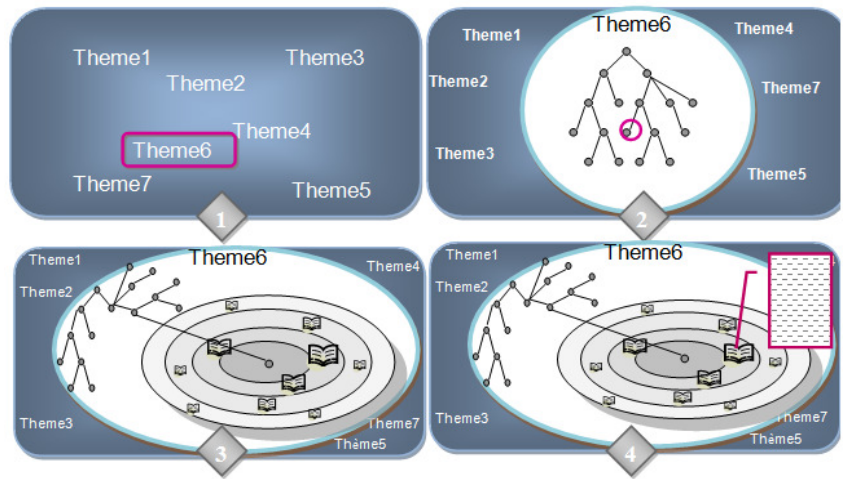

Figure 2. Navigation guided by the domain ontology (1) overview of the general themes, (2) graphic representation of the concept hierarchy of the selected theme, (3) visualization of the document indexed by the selected concept with their respective degree of pertinence, (4) visualization of the semantic annotation summary of the selected document

The visualization example, presented by Figure 3, illustrates the navigation path which a user can make to access to documents indexed by the concepts "Multi-agent System". The Figure 3.a corresponds to a view of the domain ontology representation centered on three themes: *security*, *artificial intelligence* and *information system*.

In this arborescence the user navigates to localize his interest center. In this example, the user chooses initially the theme "*Artificial Intelligence*", then he chooses the subtheme "*Application and expert systems*". After consulting the map the user selects the concept "*Multi-Agent System*". A view containing titles and relevance degrees of documents annotated by this concept is shown allowing the user to make a global idea about the documents indexed by this concept and their respective pertinence (Figure 3.b). The user can display a detailed description of every document before downloading it or visualizing it in the full text.

Figure 4 illustrates an example of a document description selected by a user. The document description corresponds to the descriptive annotations, the key concepts, the cooccurrence hypergraph and the thematic composition of the document (major theme, minor themes). This figure shows that the document deals with three themes: mainly "*Artificial Intelligence: And Expert Systems application*" who is considered as the major theme and "*Security: Cryptography*" and "*Security: Network Security*" who are considered as minor themes.

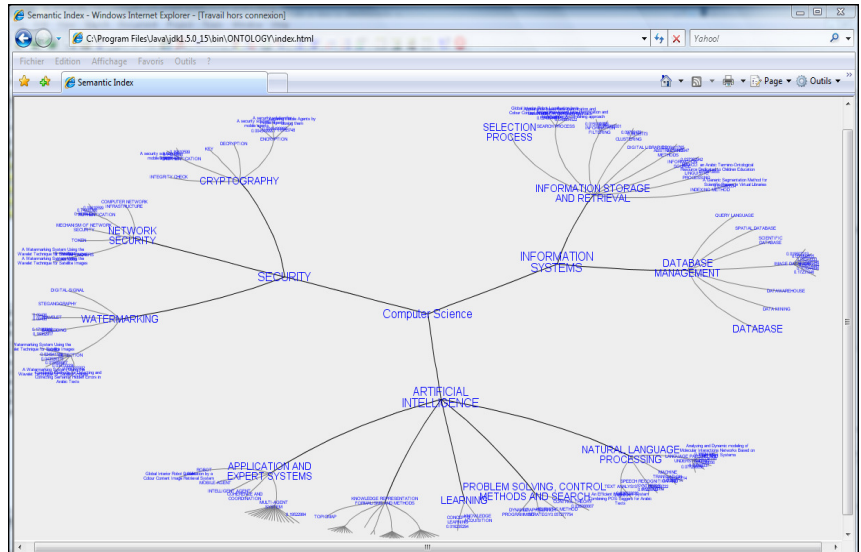

(a)

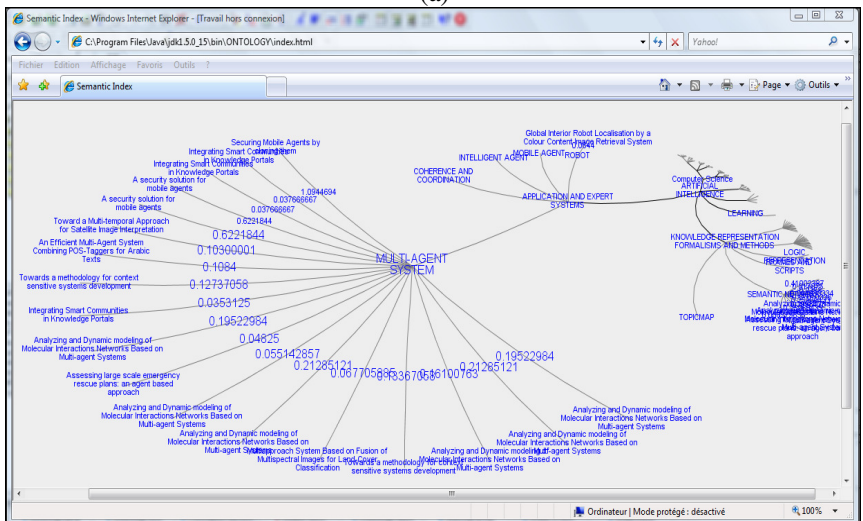

(b)

Figure 3. Example of navigation path

## 3.2. Association relation based navigation

When navigating in the domain ontology, the user can focus his attention on a concept and wishes to know what are the concepts associated to it (Figure 5). The analysis of the conceptual association relations in the corpus allows answering this kind of needs. Our idea is to build for every concept an association graph allowing users to discover the association relations of their interest concept and to visualize documents relative to an association of their choice. For the identification of the conceptual association graphs we are based on the construction and the analysis of the cooccurrence networks [18].

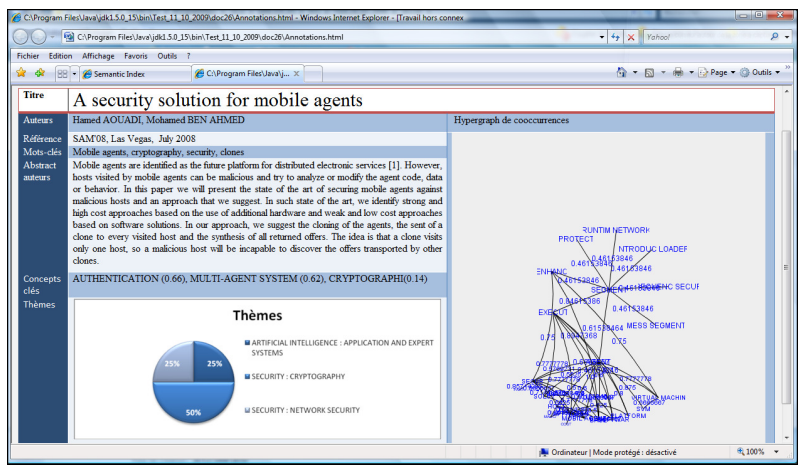

Figure 4. Example of document selected in the first path

So, for every concept of the ontology we determine the set of the concepts with which it is associated by a cooccurrence relation. We measure the degree of association of every relation according to the number of documents in which both concepts collocate. The analysis of the cooccurrence relations between concepts allows to index documents by conceptual associations.

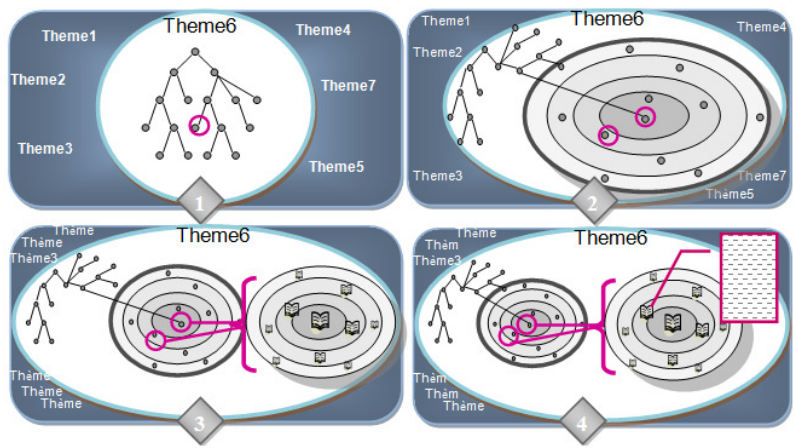

Figure 5. Association Relation based Navigation (1) graphic representation of the domain ontology, (2) visualization of the association relations of the selected concept, (3) visualization of the documents indexed by the selected association relation, (4) visualization of the semantic annotation summary of the selected document

Figure 6 shows an example of concept association hypergraph concerning the concept "*Multi-Agent System*". The central node represents the concept of interest. The first level of nodes represents the set of concepts associated to the central concept. The label of an edge between the central concept and another concept represents the association degree between the two concepts. From this figure, we can note for example that the concept "*Multi-Agent System*" is associated with the concept "*Semantic Network*" with an association degree equal to 0.36 and to the concept "*Authentication*" with an association degree equal to 0.13.

The second level of nodes represents documents indexed by these conceptual associations. The label of an edge which connects a *document node* and an *associated concept node* represents the relevance degree of the document with regard to both concepts (*associated concept* and *central concept*). For example, Figure 6.b shows that three documents are indexed by the association relation between the two concepts "*Multi-Agent System" and "Authentication*". The first document entitled "*A security solution for mobile agents*" has a relevance degree equal to 0.642. The second document entitled "*Securing Mobile Agents*" has a relevance degree equal to 0.682. The third document entitled "*Securing Mobile Agents by cloning them*" has a relevance degree equal to 0.645.

The main interest of integrating conceptual association relations in the visualization process is to allow users to discover information related to their initial interest center what contributes to enlarge their domain knowledge. Besides, the visualization of association relations allows reflecting the real context in which concepts are evoked in documents. So users could refine their search according to the conceptual associations which are relevant to them (filter documents) and to discover new knowledge.

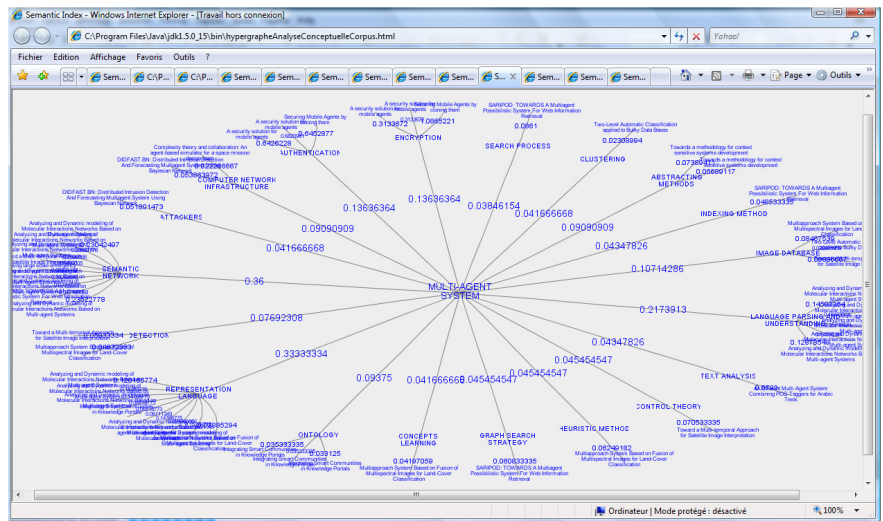

(a)

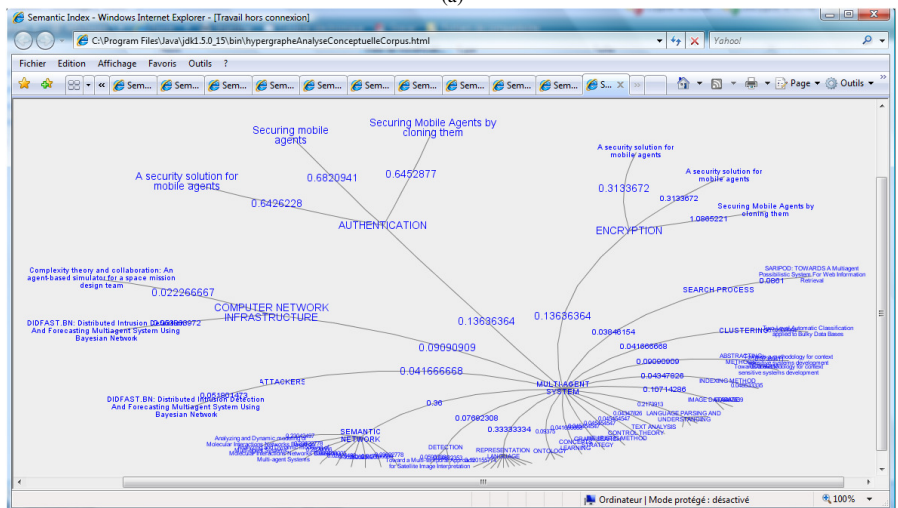

(b)

Figure 6. Navigation in the association Hypergraphe of the concept « Multi-Agent System »

## 4. CONCLUSION

The evaluation of information visualization is a very problematic task [24] [25]. Several challenges could rise when researchers conduct an information visualization evaluation. These challenges can be related to many factors: the context of use, participant gathering, data collection, existence of evaluation environment (standard, reference tool for comparison, etc.).
As first future work, we intend to focus our attention on studying the existing method and metric of evaluation of information visualization and semantic maps in order to evaluate our approach of semantic navigation.

One of the biggest challenges of the visualization conception is that there is no strategy of "ideal" visualization; the conception is always specific to the application. Different systems are efficient for users having different backgrounds and needs (expert or novice, scientist or general information). A universal model is difficult to be generalized.
As another perspective, we plan to construct a toolbox allowing users to select interactively the visualization paradigm to be used in their maps and to make conversion between visualization paradigms if they are not satisfied.

## REFERENCES

[1] W. Chung, H. Chen, J.F. Nunamaker, Business Intelligence Explorer : A Knowledge Map Framework for Discovering Business Intelligence on the Web, Proceedings of the 36th Hawaii International Conference on System Sciences (HICSS'03), 2003

[2] S.K. Card, J.D. Mackinlay, B. Sheinderman, Information Visualisation, Readings in information visualization: using vision to think, Morgan Kaufmann Publishers Inc., 1999, p.1-34.

[3] C. Tricot, Cartographie sémantique, des connaissances à la carte, thèse de doctorat, Université se Savoie, 2006.

[4] R. Rao, S.K. Card, The Table Lens: Merging Graphical and Symbolic Representations in an Interactive Focus+Context Visualization for Tabular Information, in Beth, A., Susan, D. et Judith, S.O. (eds.), ACM Press, 1994, p.318-322.

[5] F. van Ham, J.J. van Wijk, Beamtrees: compact visualization of large hierarchies, Information Visualization, 2, 1, 2003, p.31-39.

[6] S., Havre, B., Hetzler, L., Nowell, ThemeRiverTM*: In Search of Trends, Patterns, and Relationships, In IEEE Transactions on Visualization and Computer Graphics, volume 8, pages 9–20. IEEE Computer Society Press, 2002

[7] J.V. Carlis, J.A. Konstan. Interactive visualization of serial periodic data. In Proceedings of the 11th Annual ACM Symposium on User Interface Software and Technology, pages 29–38. ACM Press, 1998.

[8] J.A. Wise, J.J. Thomas, K. Pennock, D. Lantrip, M. Pottier, A. Schur, V. Crow, Visualizing the non-visual: spatial analysis and interaction with information from text documents, Proceedings of the 1995 IEEE Symposium on Information Visualization, Atlanta, Georgia, IEEE Computer Society, 1995

[9] D.F. Jerding, J.T. Stasko, The information mural: a technique for displaying and navigating large information spaces, Proceedings of the 1995 IEEE Symposium on Information Visualization, Atlanta, Georgia, IEEE Computer Society, 1995

[10] G.G. Robertson, J.D. Mackinlay, The document lens, Proceedings of the 6th annual ACM symposium on User interface software and technology, Atlanta, Georgia, United States, ACM Press, 1993

[11] J.D. Mackinlay, G.G. Robertson, S.K. Card, The perspective wall: detail and context smoothly integrated, Proceedings of the SIGCHI conference on Human factors in computing systems: Reaching through technology, New Orleans, Louisiana, United States, ACM Press, 1991

[12] B.B. Bederson, Fisheye menus, Proceedings of the 13th annual ACM symposium on User interface software and technology, San Diego, California, United States, ACM Press, 2000

[13] M. Hascoët-Zizi, N. Pediotakis, Visual Relevance Analysis. Proc. ACM Conference on Digital Libraries (DL'96). 1996, pp. 54-62.

[14] L. Bartram, A. Hot, J. Dill, F. Henigman, The Continuous Zoom: A Constrained Fisheye Technique for Viewing and Navigating Large Information Spaces, Proc. ACM Symposium on User Interface Software and Technology (UIST'95), 1995, pp. 207-215.

[15] M. Hascoët, M. Beaudouin-Lafon, Visualisation Interactive d'Information, I3: Information, Interaction, Intelligence, Vol. 1, n° 1, 2001, p.77-108.

[16] J. Zhang, Visualization for Information Retrieval, ISBN: 978-3-540-75147-2 e-ISBN: 978-3-540-75148-9, Springer-Verlag Berlin Heidelberg, 2008

[17] Férihane Kboubi, Anja Habacha Chabi, Mohamed BenAhmed, Semantic Cartography in Information Retrieval Systems, International Journal of Advanced Science and Technology, accepted in Press.

[18] F. Kboubi, A. Habacha, M. Ben Ahmed, Mining Term Association Rules for Conceptual Annotation of Textual Document, International Journal of Intelligent Information Technology Application IJIITA, 2009.

[19] F. Kboubi, A. Habacha, M. Ben Ahmed, L'exploitation des relations d'association de termes pour l'enrichissement de l'indexation de documents textuels, JADT 2010

[20] Anja Habacha Chabi, Ferihane Kboubi, Mohamed Ben Ahmed, Thematic Analysis and Visualization of Textual Corpus, International Journal of Computer Science & Engineering Survey (IJCSES), November 2011, Volume 2 Number 4

[21] N. Gershon, W. Page, What Storytelling can do for information visualization, In : Communication of ACM, Vol.44, n°8, 2001.

[22] J. Nielsen, Designing the Web Usability. Indianapolis: New Riders Publishing, 2000.

[23] K. Cassidy, A. Walsh, and B. Coghlan, Using Hyperbolic Geometry for Visualization of Concept Spaces for Adaptive eLearning. A3H: 1st Inter. Workshop on Authoring of Adaptive & Adaptable Hypermedia, June 20, 2006, Dublin, Ireland .

[24] J. Venna, J. Peltonen, K. Nybo, H. Aidos, S. Kaski, Information Retrieval Perspective to Nonlinear Dimensionality Reduction for Data Visualization, Journal of Machine Learning Research 11, 2010, pp 451-490

[25] M. Sedlmair, P. Isenberg, D. Baur, A. Butz, Evaluating Information Visualization in Large Companies: Challenges, Experiences and Recommendations, Proceedings of the CHI Workshop Beyond Time and Errors: Novel Evaluation Methods for Information Visualization (BELIV). Atlanta, USA, 2010.